\documentstyle[10pt,times,amssymb,a4,epsf]{article}
\renewcommand{\P} {{\mathbb P}}
\newcommand{\C}{{\mathbb C}}
\newcommand{\Z}{{\mathbb Z}}
\def\&#1{^{[{#1}]}}
\begin{document}
\bibliographystyle{perso}

\begin{titlepage}
\null \vskip -0.6cm

\null \hfill {PAR--LPTHE 98--22}

\vskip 1.4truecm
\begin{center}
\obeylines

     {\Large
     	Algebraic entropy.}

\vskip 1cm
M.P. Bellon$^a$ and C-M. Viallet$^{a,b}$
\em $^a$ Laboratoire de Physique Th\'eorique et Hautes Energies
Unit\'e Mixte de Recherche 7589, CNRS et Universit\'es Paris VI et Paris VII.
$^b$  CERN, Division TH
\end{center}

\vskip 2cm
\noindent{\bf Abstract}: 
For any discrete time dynamical system with a rational evolution, we define an 
entropy, which is a global index of complexity for the evolution map.  We
analyze its basic properties and its relations to the singularities and the
irreversibility of the map.  We indicate how it can be exactly calculated.
\vfill

\begin{center}

\medskip\hrule \medskip
\obeylines
Postal address: %
Laboratoire de Physique Th\'eorique et des Hautes Energies
Universit\'e Pierre et Marie Curie, bo\^\i te postale 126.
4 Place Jussieu/ F--75252 PARIS Cedex 05

\end{center}
\end{titlepage}

\section{Introduction}

Exploring the behaviour of dynamical systems is an old
subject of mechanics~\cite{Li55,Po92}. 
Turning to discrete systems has triggered a huge activity 
and the notions of sensitivity to initial conditions, numerical
(in)-stability, Lyapunov exponents and various entropies remained at the
core of the subject (see for example~\cite{EcRu85}).

We describe here the construction of a characterizing number
associated to discrete systems having a rational evolution (the state at time
$t+1$ is expressible rationally in terms of the state at time $t$): it
is  defined in an algebraic way and we call it the {\em algebraic 
entropy} of the map.
It is linked to global properties of the evolution map, which  usually is
not everywhere invertible.  It is not attached to any particular domain of
initial conditions and reflects its asymptotic behaviour.  Its definition
moreover does not require the existence of any particular object like
an ergodic measure.

In previous works~\cite{FaVi93,BoMaRo94,HiVi97,Ve92},  a link has been
observed between the dynamical complexity~\cite{Ar90} and the degree of the composed map. 
The naive composition of $n$ degree-$d$ maps is of 
degree $d^n$, but common factors 
can be eliminated without any change to the map on generic points.
This lowers the degree of the iterates.  For maps admitting invariants, the
growth of the degree was observed to be polynomial, while the generic
growth is exponential.  

We first define the algebraic entropy of a
map from the growth of the degrees of its iterates, and give some of its
fundamental properties.  From the enumeration of the degrees of the first
iterates, it is possible to infer the generating function and extract the
exact value of the algebraic entropy, {\em even for systems with a large number
of degrees of freedom}.  The reason underlying this
calculability is the existence of a finite recurrence relation between the
degrees.

After reviewing basic properties of birational maps and of their
singularities, we prove such recurrences for specific families.  The proof
relies on the analysis of the singularities.  

We describe the relations between the factorization process 
governing the growth of the degrees of the iterates and the geometry of
the singularities of the evolution.  This put a new light on the analysis
of~\cite{GrRaPa91,RaGrHi91b,GrNiRa96}.  See also~\cite{HiVi97b}.

\section{Algebraic entropy}
\subsection{Definition}

The primary notion we use is the degree of a rational map.  
In order to assign a well defined degree to a map, we require that all
the components of the map are reduced to a common denominator of the
smallest possible degree. 
The maximum degree of the common
denominator and the various numerators is called the degree of the rational
maps and it is the common degree of the homogeneous polynomials describing
the map in {\em projective space}.  From this definition we obtain the two 
basic properties:
\begin{itemize}
\item The degree is invariant by projective transformations of the source and
image spaces.
\item The degree of the composition of two maps is bounded by the 
product of the degrees of the maps.
\end{itemize}
From now on, rational maps will always be defined by 
homogeneous polynomials acting on homogeneous coordinates of the projective
completion of affine space. 
When calculating the composition of two maps, common factors may appear which
lower the degree of the resulting map.  We then define a reduced
composition $\phi_2\times\phi_1$ of $\phi_1$ and $\phi_2$ by:
\begin{equation}
\label{reduced}
	\phi_2 \circ \phi_1 = m(\phi_2,\phi_1) \cdot (\phi_2\times\phi_1).
\end{equation}
We denote by $\phi\&n$ the ``true"
$n$-th iterate of a map $\phi$, once all factors have been removed.

For a transformation $\phi$, we can define the sequence $d_n$ of the degrees
of the successive iterates $\phi\&n$ of $\phi$.

\proclaim Defining proposition.
The sequence $1/n \log d_n$ always admits a limit as $n\to\infty$.
By definition, we call this limit the algebraic entropy of the map $\phi$.

The proof is straightforward and is a consequence of the inequality
$ d_{n_1+n_2} \leq d_{n_1} d_{n_2}$.

The algebraic entropy is independent of the particular representation of 
the rational map $\phi$.  Indeed, if we take the conjugation of $\phi$ by
some birational transformation $\psi$, $\phi' = \psi^{-1} \times \phi \times
\psi$, the degree $d'_n$ of $\phi'^{[n]}$ will satisfy $d'_n\leq k d_n$ for
some constant $k$ depending on the degree of $\psi$.  A similar inequality
can be obtained when writing $\phi = \psi \phi' \psi^{-1}$.  In other words,
the entropy is a {\em birational invariant} associated to $\phi$.

This quantity can be rather easily computed by taking the images of an
arbitrary line.  The convergence to the asymptotic behavior is quite fast and
can be obtained from the first iterates for which the degree can be
exactly calculated.  

The growth of $d_n$ measures the complexity of the evolution, since 
$d_n$ is the number of intersections of the $n$th image of a generic line
with a fixed hyperplane.   It is related to the complexity introduced by 
Arnol'd~\cite{Ar90}, with the difference that we are not dealing with
homeomorphisms.  

The algebraic entropy also has an analytic interpretation.  An invariant
K\"ahler metric exists on the complex projective space $\P^n$ and the
volume of a $k$-dimensional algebraic variety is given by the integral
of the $k$th power of the K\"ahler form.  This volume is proportional to
the degree of the variety~\cite{Mu95}. The area of the image by
$\phi^{[n]}$ of a complex line can be expressed as the integral of the
squared modulus of the differential of $\phi^{[n]}$.  It is proportional
to $d_n$ by the above argument.  The algebraic entropy can then be
viewed as an averaged exponent: it does not depend on the choice of a
starting point and  it has the advantage of being of a global nature.

The definition of the algebraic entropy can be generalized to sequences of
maps $(\phi_k)_k$ such that the degree of $\phi_k$ is bounded.
We define $\phi\&n$ to be the regularized map $\phi_n \times \ldots \times
\phi_2 \times \phi_1$.  This allows the extension to non-autonomous
iterations
and to maps which are the product of elementary steps, in which case the
sequence $(\phi_k)_k$ is periodic.

In the cases where $d_n$ has a polynomial dependance on $n$, the algebraic
entropy is zero, but we can make use of a new invariant, the degree of $d_n$.
As the algebraic entropy, it is a birational invariant.

\subsection{Entropy of the H\'enon map}

For a simple confrontation of the algebraic entropy with more usual approaches, 
let us consider the much studied H\'enon map~\cite{He76}.  Since it is a
polynomial map,
it is usually considered as having no singularities.  This is a misconception:
using projective space shows that singularities exist and are located on 
the line at infinity.
\begin{eqnarray}
\label{henon}
	t &\rightarrow& t^2,\\
	x &\rightarrow& t^2 + t y  - a x^2, \\
	y &\rightarrow& b t x.
\end{eqnarray}
Here $t$ is the third homogeneous coordinate.  We immediately see
on this expression that the line at infinity $t=0$ is sent to the
point with homogeneous coordinates $(t,x,y)=(0,1,0)$.  This point 
is still on the line $t=0$, so it is a fixed point of the transformation.
It will therefore never be mapped to $(0,0,0)$ and there cannot be
any factorization.  The $n$th iterate of this map is of
degree $2^n$ and the algebraic entropy is $\log(2)$.
The remarkable thing is that this number is independent of the parameters
$a$ and $b$, contrarily to usual dynamical exponents.

\section{Birational maps.}
Among rational maps, we mainly use birational ones.  They are almost
everywhere invertible and are therefore quite appropriate for 
modeling systems possessing a certain amount of reversibility.

\subsection{A little bit of algebraic geometry.}
\label{basic}

Rational relations between two algebraic sets $X$ and $Y$ are relations 
with a graph $Z$ which is an algebraic subset of $X\times Y$.  It would be
to restrictive to impose that this define a map from $X$ to $Y$.  In fact, 
the only algebraic graphs defining maps on the whole space $\P^n$ define linear
transformations.  One therefore only requires that a rational map is one
to one on the complement of an algebraic variety, that is a Zariski open
set.  A birational map defines a bijection from an open subset $X_0$ of $X$
to an open subset $Y_0$ of $Y$. 

If we call $p_1$ and $p_2$ the projection
on the components of the Cartesian products restricted to $Z$, the point
$x$ will correspond to $p_2( p_1^{-1}(x))$.  When this subset of $Y$ is not 
reduced to a point,  $x$ is by definition in the singular locus
of $Z$.  If we solve for the homogeneous
coordinates of the image point, we get homogeneous polynomials in the 
coordinates of $x$.  We therefore get a map defined in $\C^{n+1}$.  
Homogeneity makes it compatible with the scale relation defining projective
space.  Some vector lines in $\C^{n+1}$ however 
are identically mapped to zero: they are projective 
points without definite images.  The set of these points is exactly the
singular locus. 


If $\phi$ is the homogeneous polynomial representation of a rational map
and $P$ is a homogeneous polynomial in
$(n+1)$ variables, we denote $\phi^*\, P$ the pull-back of $P$ by $\phi$.
It is simply obtained by the composition 
$P \circ \phi$.  The hypersurface of equation $\phi^*\, P = 0$ is the image
by $\phi^{-1}$ of the hypersurface $P=0$.
If $x_j$ is one of the homogeneous coordinates of $\P^n$, $\phi^*\, \,x_j$ is
simply the index $j$ component of the polynomial function $\phi$.
Homogeneous polynomials do not define functions on $\P^n$, but sections of a
line bundle which only depends on the homogeneity degree.

\subsection{Two examples} \label{example}
Let us describe two elementary examples of birational maps.  
The first one is the generalized Hadamard inverse in $\P^n$.
Take two copies of $\P^n$ with homogeneous coordinates $(x_0,x_1,\ldots,x_n)$
and $(y_0,y_1,\ldots,y_n)$ and define $Z$ by
the $n$ equations in $\P^n\times\P^n$:
\begin{equation}\label{cremona}
	x_i y_i = x_0 y_0,\quad\mbox{i=1,\ldots,n}
\end{equation}
On the subset where all the $x_i$ are different of zero, we can
use affine coordinates by fixing $x_0=1$ and $Z$ defines the map 
\begin{equation}
	(x_1,\ldots,x_n) \to (1/x_1,\ldots,1/x_n).
\end{equation}
If any of the $x_i$ is zero, then all the products $x_jy_j$ must be
zero.  Let $J$ be the set of indices for which $x_i$ is zero.
$Z$ induces a correspondence between $\{x\}$ and the linear space
$\bigcap_{i\in [0,\ldots,n] - J} H_i$ ($H_i$ is the hyperplane $y_i=0$).
Instead of equations~(\ref{cremona}), it is often more convenient 
to give a functional definition of this correspondence.  A polynomial
definition is:
\begin{equation}
	y_i = \prod_{j\not=i} x_i.
\end{equation}
The $y_i$'s are polynomial functions of the
$x_i$'s of degree  $n-1$ and they satisfy equations~(\ref{cremona}).
But no formula can give the proper relationship for singular points.
For these, at least two
of the $x_i$'s are zero and therefore all the $y_i$'s vanish. 

The second example is given in two dimensions by
\begin{eqnarray} \label{ratio}
	x &\to& x, \\
	y &\to& f(x) - y,
\end{eqnarray}
with $f(x)$ any rational function of $x$.  Here again, we can give
a homogeneous polynomial formulation.  We will have a third variable
$t$ which will be multiplied by the denominator of $f(x)$.

These two transformations give rise to interesting evolution maps when combined
with simple linear transformations.  The exchange of the two variables
$x$ and $y$ combined with~(\ref{ratio}) gives a family of transformation
which contains for suitable $f$ discrete versions of some Painlev\'e
equations~\cite{GrNiRa96}.
The transformation~(\ref{cremona}) and its conjugation by the Fourier
transformation yields a birational transformation which appears naturally
as a symmetry of the $(n+1)$-state chiral Potts model~\cite{BeMaVi91c}.

\subsection{Singularities}
The singular points of a birational map are the vector lines of $\C^{n+1}$
which are sent to the origin $(0,0,\ldots,0)$.  This singular set is of
codimension at least 2.  In fact, if there was an algebraic set of
codimension~1 sent to the origin, the equation of this set could 
be factored out of all the components of the image, allowing a reduced
description of the map without this singularity.

There is a bigger set where the map is not bijective.  Let $\phi$  be a
birational map
and $\psi$ be its inverse.
Than the composition $\psi \circ\phi$ of their representations as
polynomial maps in $\C^{n+1}$ is a map of degree $d^2$.  It is
however equivalent to the identity, so that each of the components
of the image are of the form $K_\phi x_i$, where $K_\phi$ is a homogeneous
polynomial of degree $d^2-1$.  The set of zeroes of $K_\phi$, $V(K_\phi)$, is
a set where the composition $\psi\circ \phi$ is a priori not defined and 
it plays a fundamental role. 

$K_\phi$ is an example of a multiplier.  When composing two birational
maps $\phi_1$ and $\phi_2$, a common factor $m(\phi_2,\phi_1)$ may appear
in the components of $\phi_2\circ\phi_1$.  In the case of inverse birational transformations, $\psi \times \phi$ 
is the identity and $m(\psi,\phi)$ is $K_\phi$.

A fundamental property of $m(\phi_2,\phi_1)$ is that it cannot vanish
out of $V(K_{\phi_1})$.  Otherwise $\phi_1$ would map an open subset of
the set of zeros of $m(\phi_2,\phi_1)$ to a codimension~1
set where $\phi_2$ is singular, since $\phi_1$ is a diffeomorphism
outside of $V(K_{\phi_1})$.  This gives us a contradiction since the singular
set of rational maps are of codimension at least 2.  Determining the 
multiplying factor amount to determining the exponents of the different
irreducible components of $K_{\phi_1}$ in $m(\phi_2,\phi_1)$.

In fact we obtain a definition of the map on a number of apparently singular
hypersurfaces, which is a natural continuous extension of the map.

\subsection{The meaning of factorization}

Consider the successive iterates $\phi\&n$ of a birational map $\phi$. 
Suppose we have the following pattern of factorization:
\begin{eqnarray}
	\phi \circ \phi &=& \phi\times\phi = \phi\&2, \label{2}\\
	\phi \circ \phi\&2 &=& \phi\times\phi\&2 = \phi\&3, \label{3}\\
	\phi \circ \phi\&3 &=&  \kappa \cdot\phi\&4, \label{4}
\end{eqnarray}
with $\kappa$ different from 1.  Equation~(\ref{4}) means that the variety
$\kappa=0$ is sent to singular points of $\phi$ by $\phi\&3$.  In other
words, $\kappa=0$ is blown down to some variety of codimension higher than
one by $\phi$.  The latter is non singular for the action of $\phi$
and $\phi\&2$ but is eventually  blown up by $\phi\&3$.

Two situations may occur: it may happen that the image by $\phi\&4$ of the
variety
$\kappa=0$ is again of codimension 1 and we have a self-regularization of the
map.  Such a situation was called singularity confinement
in~\cite{GrRaPa91,RaGrHi91b}.  We would rather call it {\em resolution of 
singularities}.  Reversibility is recovered on the singular
set of $\phi$ after a finite number of time steps.  The other possibility is
that the image
of the variety $\kappa=0$ by $\phi\&4$ remains of codimension larger than
one, a situation depicted in figure~1.

\begin{center}
\newdimen\figwidth
\figwidth = \textwidth
\advance \figwidth  by -2 cm
\epsfxsize=\figwidth
\mbox{}\epsfbox{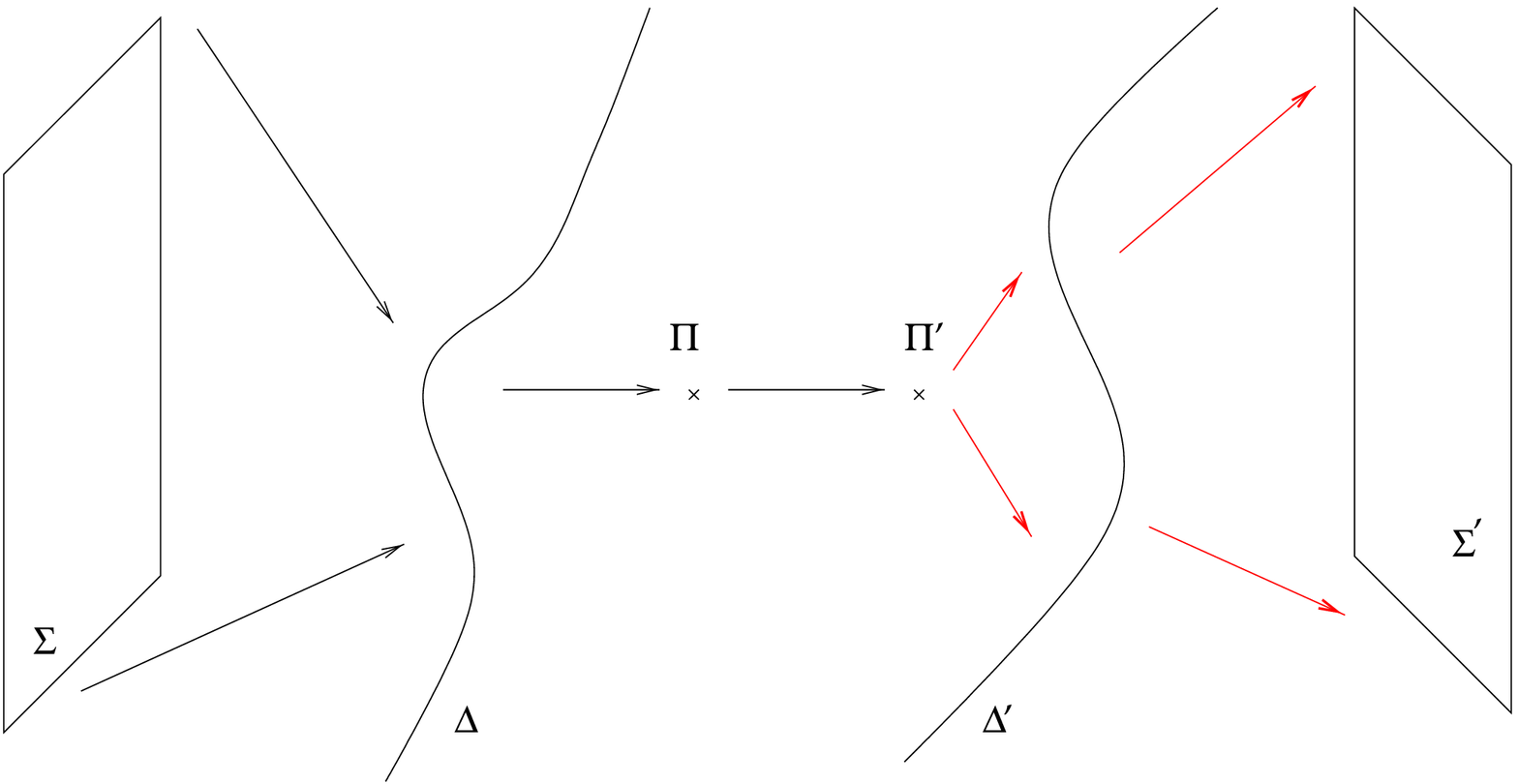}

Figure 1: A possible blow-down blow-up scheme in $\P^3$.
\end{center}

In the scheme of figure~1, the equation of $\Sigma$ is $\kappa=0$, and the
factor $\kappa$ appears anew in $\phi\circ\phi\&4$.  The fifth iterate
$\phi\&5$ is regular on $\Sigma$.

The drop of the degree of the iterates is due to the presence of
singularities on the successive images of a generic surface under the repeated
action of $\phi$.  In other words, these images are less and less generic.

\section{Recurrence relations for the degree}

One of the basic properties of the sequence of degrees is that it seemingly
always verifies a finite linear recurrence relation with integer coefficients. 
If this is true, the algebraic entropy is the logarithm of an algebraic number.  

\subsection{A simple case in $\P^2$}\label{z7}
Consider the map $\phi = \phi_2 \phi_1 $ with $\phi_1$ and $\phi_2$ given by:
\begin{eqnarray*}
\phi_1:
&&\left[ 
\matrix{x'=  x^{2}+2\,yx+2\,zx-y^{2}-z^{2}-3\,yz \cr
        y'= 2\,z^{2}-yx-y^{2} \hfill\cr z'= 2\,y^{2}-zx-z^{2}\hfill } 
	 \right. \\
	 \phi_2: &&\left[ \matrix {x'=  y z \cr y'=  x z \cr z'= x y }
 \right.
\end{eqnarray*}
It was used as an example of
chaotic behavior in~\cite{BeMaVi91c} and its singularities have been studied
in~\cite{FaVi93}, where the first few elements of the sequence $d_n$ were given:
\begin{equation}
	1,2,4,7,12,20,33,54,\ldots
\end{equation}
This sequence can be coded in the generating function:
\begin{equation}	\label{generat}
	g(x) = {1\over1-2x+x^3}.
\end{equation}
The rationality of the generating function is equivalent to the existence
of a finite linear recurrence relation for the degrees.  The determination of
the entropy is straightforward once the recurrence relation is known.

The iterated map is a product $\phi_2 \circ \phi_1$ of two linearly related
involutions $\phi_1$ and $\phi_2$ of degree 2.  It is useful in this case to
look at the sequence of iterates of $\phi$ as the sequence built from 
$(\phi_1,\phi_2,\phi_1,\phi_2,\ldots)$.

The possible factorizations come from the fact
that the line at infinity ($t=0$) is sent by $\phi_1$ to a fixed point of
$\phi_2$ and reciprocally.  When calculating $\phi_1\times\phi_2 \times \phi_1$,  
$t$ will appear as a factor, since $\phi_2 \times \phi_1$ send this line
to a singular point of $\phi_1$.  

We want to know the degree of the factor
$m(\phi^{[n]},\phi)$ or $m(\phi,\phi^{[n]})$.  
The former factor can only contain the factor $t$, but the
exponent is not readily known, so we rather examine $m(\phi,\phi^{[n]})$. 
We have to determine the curve which is sent by $\phi\&{n-2}$ to the
line $t=0$.  This is just the first component of the polynomial expression of
$\phi\&{n-2}$ and can be written $(\phi\&{n-2})^*\, t$.
It has degree $d_{n-2}$.    In two steps, the line $t=0$ is mapped to
a singular point of the following $\phi_i$.  The curve with equation
$(\phi\&{n})^*\, t$ is therefore mapped to a singular point by $\phi\&{n+2}$
and its equation can be factorized in the calculation of $\phi\&{n+3}$.
This gives:
\begin{equation}
	(\phi\&{n-2})^*\, t\cdot \phi\&{n+1} = \phi \circ \phi\&{n},
\end{equation}
and consequently the following recurrence relation for $d_n$:
\begin{equation}
\label{recur1}
	d_{n+1} = 2  d_{n} - d_{n-2}.
\end{equation}
This relation proves formula~(\ref{generat}) and yields for the degrees
an exponential growth with entropy $\log{1\over2}(1+\sqrt5)$.

\subsection{Factors in the Factors}
\label{Factor}

The previous analysis is simple because the image of $t$ remains an irreducible
polynomial.  This cannot
be true in general, since the factors in $K_\phi$ generally break into pieces
under further transformation by $\phi$~\cite{BoMaRo94}.

Let us take two birational transformations $\phi_1$ and $\phi_2$ with
respective inverses $\psi_1$ and $\psi_2$ and calculate $\psi_1
\circ \phi_1 \circ \phi_2$ in two different ways:
\begin{eqnarray}
\label{phiK}
	\psi_1 \circ \phi_1 \circ \phi_2 &=& (K_{\phi_1}\cdot Id) \circ \phi_2
	\;=\; \phi_2^*\, K_{\phi_1} \cdot \phi_2	\nonumber \\
	&=& m(\phi_1,\phi_2)^{d_{\psi_1}} \cdot \psi_1\circ(\phi_1\times\phi_2).
\end{eqnarray}
Since the components of $\phi_2$ cannot have any common factor, we
deduce that $m(\phi_1,\phi_2)^{d_{\psi_1}}$ divides $\phi_2^*\, K_{\phi_1}$.

Geometrically, $m(\phi_1,\phi_2)$ is the equation of a hypersurface which
$\phi_2$ sends to singular points of $\phi_1$.  Since $K_{\phi_1}$ vanishes
on the singular points of $\phi_1$, its image $\phi_2^*\, K_{\phi_1}$ vanishes 
on the zero locus of $m(\phi_1,\phi_2)$.  

In the example of the previous section, each new factor appearing in $\phi\circ 
\phi\&n$ is the equation of a hypersurface which $\phi\&n$ sends 
to the point $(1,0,0)$. The $x$ and $y$ components of $\phi\&n$
therefore have a common factor $m(\phi,\phi\&n)$.  Consequently, the image
$\phi\&n{}^*\,K_\phi$ of $K_{\phi}=t\,x\,y$ by $\phi\&n$ contains
the expected factor $m(\phi,\phi\&n)^2$, while $\phi\&n{}^*\,t$ does not
contain this factor.

\subsection{An example in $\P^{N-1}$}

Consider the algebra of the finite group ${\Z}_N$, its generic
element $a(x) = \sum_{q=0}^{N-1} x_q \; \sigma^q $,
with $x=(x_0,x_1,\ldots, x_{N-1})$ and $\sigma$ the generator of ${\Z}_N$.

The algebra has two homomorphic products:
\begin{eqnarray}
a( x\circ y)& = & a(x) \circ a(y), \\
a( x \cdot y )& = & \sum_{q=0}^{N-1} x_q y_q  \; \sigma^q .
\end{eqnarray}
The product $\circ$ just comes from the product in ${\Z}_N$, and
verifies $\sigma^p \circ \sigma^q = \sigma^{(p+q)}$, while $\sigma^p
\cdot \sigma^q = \delta^p_q \sigma^p$.  In terms of cyclic matrices, these
two products respectively correspond to the matrix product and the element
by element (Hadamard) product.  The homomorphism between these two
products is realized by the Fourier transform.  

$\phi_1$ and $\phi_2$ will be the two inverses constructed from these 
products.  The components $(x_0,x_1,\ldots, x_{N-1})$ of $x$ are the natural
coordinates of projective space and $\phi_1$ and $\phi_2$ are involutions
of degree $N-1$.  $K_{\phi_1}$ and $K_{\phi_2}$ are products of linear
factors.  These linear factors are the equations of hyperplanes which are
sent by the corresponding $\phi_i$ into points which we call maximally
singular.

The important fact is that these maximally singular points are permuted
by the other involution.  As an example, the maximally singular points
of the Hadamard inverse are of the form $\sigma^q$, i.e., with only one
non zero component.  The matrix inverse permutes such points by $\sigma^q \to
\sigma^{N-q}$.

If $p$ has one vanishing component, say $x_i$, then $\phi\&2(p)$
will have all its coordinates vanishing except $x_{N-i}$.  It follows that
$x_i$ is a common factor to all these coordinates.  
The $j$-th coordinate of $\phi\&2(p)$ can be written\footnote{
As in section~\ref{z7}, we write $\phi^n$ (resp.\ $\phi\&n$) for the
composition of alternatively the two inverses (resp.\ the reduced
composition).}:
\begin{equation}
\label{fact1}
	\phi^2(p)_j = x_j\&2\prod_{i\neq N-j} x_i.
\end{equation}
The Hadamard inverse is easily calculated on such an expression.
The coordinates of $\phi^3 (p)$ are given by:
\begin{equation}\label{pas3}
	\phi^3(p)_j = \bigl(\prod_{i\neq j} x_i\&2\bigr)
		x_{N-j} \bigl(\prod_{i=0\ldots N-1} x_i \bigr)^{N-2}.
\end{equation}
The common factor is simply $K_{\phi_1}$ and this suggest that 
$\phi\&3$ is a local diffeomorphism on the zeroes of $K_{\phi_1}$. 

We now want to determine the structure of the components of $\phi\&n$
for any $n$.  The situation for $n$ odd and $n$ even will be 
similar, since the conjugation by the Fourier transform exchanges the two
inverses.  
From the
expression~(\ref{fact1}) of $\phi^2(p)$, we see that this point is a generic
element of the plane $x_j=0$ if and only if $x_j\&2=0$. 
We define polynomials $x_j\&{n+2}$ generalizing
the $x_j\&2$'s appearing in~(\ref{fact1}) such that\footnote{In this formula,
the coordinates are different according to the parity of $n$.  They are
always such that the following $\phi$ is the Hadamard inverse in those
coordinates.  }:
\begin{equation}
\label{fact2}
	\phi\&{n+2}(p)_j = x_j\&{n+2}\prod_{i\neq N-j} x_i\&n.
\end{equation}
If $g_n$ is the degree of the $x_j\&n$'s, then equation~(\ref{fact2}) yields:
\begin{equation}
	d_n = g_n + (N-1) g_{n-2}.
\end{equation}
The generalization of~(\ref{pas3}) gives: 
\begin{equation}
m(\phi,\phi\&n) = \Bigl(\prod_j x_i\&{n-2}\Bigr) ^{N-2}.
\end{equation}
The factor is therefore of degree $N(N-2) g_n$.  Finally:
\begin{eqnarray}
	g_{n+1} &=& d _{n+1} - (N-1) g_{n-1}
	= (N-1)d_n - N(N-2) g_{n-2} -(N-1)g_{n-1} \nonumber \\
	&=&  (N-1) g_n - (N-1) g_{n-1} + g_{n-2}.
\end{eqnarray}
It is easy from this recurrence relation to determine that for $N=3$, 
$(g_n)$ and therefore $(d_n)$ are periodic sequences of period 6.
The sequence of the $\phi\&n$'s is known to have this periodicity.  For $N=4$,
$g_n$ is a polynomial of degree 2 in $n$, and for bigger $N$, the sequences
are growing like $\beta^n$, with $\beta$ the larger root of $x^2 -(N-2)x +1$.

\subsection{Another proof}
There is another way to prove the previous result, relating directly 
to the study of the singularities and the blow-down blow-up process.

We first need to introduce  some notations,  using homogeneous
coordinates system for $\P^{N-1}$.  The Hadamard inverse $\phi_1$
sends $(x_0,x_1,\ldots, x_{N-1})$ into $(x_0',x_1',\ldots, x_{N-1}')$,
where $x_k'= \prod_{\alpha \neq k} x_\alpha$.  
The square of $\phi_1$ is the multiplication by $K_{\phi_1}$.

Define $C$ to be the
projective linear transformation constructed from the matrix
\begin{equation}
C= \pmatrix { 	1 & 1 & 1 & \dots & 1 \cr
		1 & \omega & \omega^2 & \dots & \omega^{N-1} \cr
		\vdots & \vdots & \vdots &\ddots& \vdots \cr
		1 & \omega^{N-1} & \omega^{2(N-1)} & \dots& \omega^{(N-1)^2} },
\end{equation}
with $\omega=\exp( {2 i \pi/{N}})$. The inverse of  $C$  is its complex
conjugate $\bar{C}$.  
The involution $\phi_2$  is linearly related to $\phi_1$ by
\begin{equation}
\phi_2 = C \; \phi_1 \; \bar{C}.
\end{equation}
The product $\phi_2 \circ \phi_1$ may thus be rewritten $\rho_1 \circ
\rho_2$,  with $\rho_1= C \circ \phi_1$ and $\rho_2 = \bar{C} \circ \phi_1$.
Denote by $\psi_1 = \phi_1 \circ \bar{C}$ and $\psi_2 = \phi_1 \circ C$
the inverses of $\rho_1$ and $\rho_2 $ respectively.

The maximally singular points of the Hadamard inverse are the points
$P_i$ with $x_i$ the only non vanishing component.
They are the blow down by $\phi_1$ of the planes $\Pi_i: \{ x_i=0\}$
for $i=0,\dots,(N-1)$. They are singular points of $\rho_1$ and
$\rho_2$.  Denote by $Q_i, i=0..(N-1)$ the points $Q_i=(1,\omega^i,
\omega^{2i}, \dots, \omega^{(N-1)i})$.  
The $Q_i$'s are singular points of $\psi_1$ and $\psi_2$.

We have the following straightforward relations:
\begin{eqnarray}
\label{p2q}
C P_i = Q_i, \qquad &&\bar{C} P_i = Q_{-i}, \qquad \bar{C} Q_i= P_i=C Q_{-i},\\
\label{pi2p}
&&\phi_1(\Pi_i) = P_i.
\end{eqnarray}
The relevant singularity structure is entirely described by the two sequences:
\begin{eqnarray}
&& \Pi_i \mathop{\rightsquigarrow}^{\phi_1} P_i \mathop{\longmapsto}^C
Q_{i} \mathop{\longmapsto}^{\phi_1} Q_{-i}
\mathop{\longmapsto}^{\bar{C}} P_{-i} \mathop{\rightsquigarrow}^{\phi_1}
\Pi_{-i}, \label{singupat1} \\ 
&& \Pi_i \mathop{\rightsquigarrow}^{\phi_1} P_i
\mathop{\longmapsto}^{\bar{C}} Q_{-i} \mathop{\longmapsto}^{\phi_1}
Q_{i} \mathop{\longmapsto}^{{C}} P_{-i} \mathop{\rightsquigarrow}^{\phi_1}
\Pi_{-i}.\label{singupat2} 
\end{eqnarray}
The first squiggly line indicates blow down from hyperplane to point and
the last one indicates blow up from point to hyperplane.

Consider now a sequence $\{S_k\}$ of varieties of codimension one,
constructed by the successive action of $\rho_1$, $\rho_2$, $\rho_1$,
and so on.  Suppose the ordering is such that $\rho_1$ acts on the
$S$'s with even index and $\rho_2$ on the $S$'s with odd index.
The successive images in the sequence are supposed to be regularized 
by continuity. We denote by $d_n = d(S_n)$ the degree of the equation of $S_n$.

Denote by $\alpha_k(n)$ (resp.\ $\beta_k(n)$) the order of $P_k$
(resp.\ $Q_k$) on $S_n$. If $a$ is the running point of $\P^{N-1}$, then we have the defining relations
\begin{eqnarray}
\label{defining1}
S_{2n}( \rho_2(a))& =& S_{2n-1}(a) \cdot \prod_{u=0}^{N-1}
x_{-u}^{\beta_u(2n)}(a) ,\\
\label{defining2}
S_{2n-1}( \psi_2(a)) &=& S_{2n}(a) \cdot \prod_{v=0}^{N-1}
x_v^{\alpha_v(2n-1)} (C \;  a).
\end{eqnarray}
Using the fact that $\rho_i$ and $\psi_i$ are inverse of each other,
and relations (\ref{defining1},~\ref{defining2}), we get by evaluating
$S_{2n}(\rho_2 \psi_2 (a)) = K_{\psi_2}^{d_{2n}} \cdot
S_{2n}(a)$, the following relation on the degrees:
\begin{eqnarray}
\label{deg1}
(N-1)\; d_{2n} = \alpha_v(2n-1) + \sum_{b\neq -v} \beta_b(2n).
\end{eqnarray}
Similarly, calculating $S_{2n}( \psi_1 \rho_1 (a))$ produces
\begin{eqnarray}
\label{deg2}
(N-1)\; d_{2n} = \beta_u(2n+1) + \sum_{k\neq u} \alpha_u(2n).
\end{eqnarray}
Let $\Theta_\alpha(n) = \sum_k \alpha_k(n)$ and $\Theta_\beta(n) = \sum_k
\beta_k(n)$. Relations~(\ref{deg1},\ref{deg2}) yield
\begin{eqnarray}
\label{onetwo}
N(N-1)\; d_{2n} &=& \Theta_\alpha(2n-1) + (N-1) \;\Theta_\beta(2n),\nonumber\\
N(N-1)\; d_{2n} &=& \Theta_\beta(2n+1) + (N-1)\; \Theta_\alpha(2n).
\end{eqnarray}
From the singularity pattern~(\ref{singupat1},\ref{singupat2}), we see that 
$ \alpha_i(2n) = \beta_{-i}(2n-1)$ and $ \alpha_i(2n+1) = \beta_i(2n) $,
so that
$ \Theta_\alpha(2n) = \Theta_\beta(2n-1) $ and $ 
\Theta_\alpha(2n+1) = \Theta_\beta(2n) $.
It follows that
\begin{equation}
\Theta_\alpha(k) = \Theta_\beta(k-1).
\end{equation}
This combined with (\ref{onetwo}) yields 
\begin{eqnarray}
d_{n+3} - (N-1) d_{n+2} + (N-1) d_{n+1} - d_n,
\end{eqnarray}
which is the recurrence relation on the degrees of the iterates,
with generating function 
\begin{eqnarray}
f_q(z) = {{ 1+ z^2 (N-1) } \over { (1-z)(z^2 -z(N-2) +1) }}.
\end{eqnarray}

\subsection{Discrete Painlev\'e I}

The discrete Painlev\'e I system is given by the following transformations:
\begin{eqnarray}
\label{PI}
	x &\to& {c_n\over x} + b  - x - y, \nonumber\\
	y &\to& x,
\end{eqnarray}
where $c_n$ depends on three parameters and is given by $c_n = c + a n +d(-1)^n$. 
The transformation is just an involution of the form~(\ref{ratio}) followed
by the exchange of $x$ and $y$.  The homogeneous form is $\phi_n$ given by:
\begin{equation}
	\label{PIh}
	(t,x,y) \to ( xt, c_n t^2 + b x t + x^2 - yx, x^2).
\end{equation}
It is easy to obtain that $K_\phi$ is simply $x^3$, so that $x$ is the only
factor which can appear in $m(\psi,\phi)$.  The line $x$ is sent to the
point of coordinates $(0,1,0)$, but it is not sufficient to characterize a
possible blowing up.  In fact, at leading order in $x$, the image of points
approaching this line satisfy the equation $xy=c_nt^2$.  We therefore have to
follow the image up to second order.  $x$ remains a factor in the
successive transformations of $t$.  In $(\phi^3)^*\,x$, $x^2$ appears as a
factor.  This gives a factor of $x^6$ in the transformation of $K_\phi$ and
is the signal according to section~\ref{Factor} of the factor $x^3$ appearing 
in $m(\phi,\phi^3)$.  The $x^2$ factor is however not sufficient to guarantee
the factorization of $x^3$ in the next composition.  The factorization of 
$x^3$ depends on the relation $c_{n+3}-c_{n+2}-c_{n+1}+c_n$ which
characterizes the form of the $c_n$ given in~\cite{GrNiRa96}.

We may now establish the recurrence relation obeyed 
by the degrees.  We introduce the polynomials $x\&n$ of degree $g_n$ such
that the $x$ component of $\phi\&n$ is $x\&n(x\&{n-3})^2$.  As in the 
preceding case, the factor which will replace $x$ in the successive
factorization is $x\&n$.  The factors $x\&{n-3}$ have disappeared  in 
$\phi\&{n+1}$ and the images of $x\&{n-3}=0$ are nor singular. 
Since $d_n = g_n + 2 g_{n-3}$, we have:
\begin{equation}
	d_{n+1} = 2d_n - 3 g_{n-3} = 2 g_n +g_{n-3}.
\end{equation}
whose solution is:
\begin{eqnarray}
	g_n &=& \textstyle (1 + {1\over2}n)^2 - {1\over8}(1 - (-1)^n), \\
	d_n &=& \textstyle {3\over4} n^2 + {9\over8} - {1\over8}(-1)^n.
\end{eqnarray}
These results agree with the explicit calculations, producing the sequence
of degrees:
\begin{equation}
	1,2,4,8,13,20,28,38,49,62,76,\ldots
\end{equation}

We can also consider a slight generalization introduced in~\cite{HiVi97}.
The pole part of the transformation of $x$ is replaced by a double pole but
we do not use variable coefficients.
\begin{eqnarray}
	 x &\to& {c\over x^2} + b  - x - y, \nonumber\\
	      y &\to& x.
\end{eqnarray}
This is now a degree 3 birational map, with $K_\phi=x^3$.  It was shown that
we still have the same pattern, but with higher powers of $x$ appearing.  In 
$\phi\&3$, the $x$ component gets a $x^3$ factor and we can factorize
$x^8$ from $\phi^4$.  Defining similarly $x\&n$ such that the $x$ component
of $\phi\&n$ is $x\&n(x\&{n-3})^3$, we get the following recurrence relation
for its degree $g_n$:
\begin{equation}
	g_{n+1} - 3 g_n + 3 g_{n-2} -g_{n-3} = 0.
\end{equation}
The solution of this equation allows to recover the results
of~\cite{HiVi97}.  The algebraic entropy is given by the logarithm of the 
largest solution of $x^4-3x^3+3x-1$ which is ${1\over2}(3+\sqrt5)$, the
square of the golden ratio.

\section{Conclusion and perspectives}
We have not produced the general proof of the existence of a finite
recurrence on the degrees.  We have however shown that its origin lies
in the singularity structure of the evolution and the possible
recovering of reversibility.

In numerous examples which we will not enumerate, we have been able either
to establish recurrence
equations or to infer a generating function from the first degrees which
successfully predict the following ones.  This supports 
the following conjecture.
\proclaim Conjecture.
The generating function of the sequence of the degrees is always a rational
function with integer coefficients.

This may even be the case for rational
transformation which are not birational~\cite{AbAnBoMa98}.
The algebraic entropy is in this case the logarithm of an
algebraic number and  in the case of vanishing entropy, the sequence of the
degrees is of polynomial growth.

There is a keyword which we did not use yet: integrability.   Proving
integrability in our setting amounts to showing that the motion is a
translation on a torus.  From the numerous examples we have examined, we
believe the algebraic entropy measures a deviation from this type of
integrability.  We can actually propose the following:
\proclaim Conjecture 2.
If the birational transformation $\phi$ is equivalent to a bijection defined
on an algebraic variety $M$ deduced from $\P^n$ by a finite sequence of
blow-ups,
then the sequence of degrees of $\phi\&n$ has at most a polynomial growth.

There also is the question of the relation of the algebraic entropy to other
dynamical entropies~\cite{EcRu85}.  The fact that the algebraic maps we study
do not necessarily admit an ergodic measure precludes the definition of the
Kolmogorov-Sinai entropy in many cases.  The most natural correspondance
would be with the topological entropy, but requires more work.  We must also
stress that in any case, the algebraic entropy is a property of the map in
the complex domain.

The special properties of rational maps allow to characterize the complexity of
the dynamics from the study of a single number, the degree, and to control 
it through the study of the behaviour of the map in a small number of
singular points.


\begin{thebibliography}{10}

\bibitem{Li55}
J.~Liouville, {\em Sur l'int\'egration des \'equations diff\'erentielles de la
  {D}ynamique}.
\newblock Journal de Math\'ematiques Pures et Appliqu\'ees {\bf XX} (1855), pp.
  137--138.

\bibitem{Po92}
H.~Poincar\'e.
\newblock {\em Les m\'ethodes nouvelles de la m\'ecanique c\'eleste}.
\newblock Gauthier--Villars, Paris,  (1892).

\bibitem{EcRu85}
J.-P. Eckmann and D.~Ruelle, {\em Ergodic theory of chaos}.
\newblock Rev. Mod. Phys. {\bf 57}(3) (1985), pp. 617--656.

\bibitem{FaVi93}
G.~Falqui and C.-M. Viallet, {\em Singularity, complexity, and
  quasi--integrability of rational mappings}.
\newblock Comm.\ Math.\ Phys. {\bf 154} (1993), pp. 111--125.

\bibitem{BoMaRo94}
S.~Boukraa, J-M. Maillard, and G.~Rollet, {\em Integrable mappings and
  polynomial growth}.
\newblock Physica A {\bf 209} (1994), pp. 162--222.

\bibitem{HiVi97}
J.~Hietarinta and C.-M. Viallet.
\newblock {\em Singularity confinement and chaos in discrete systems}.
\newblock solv-int/9711014,  (1997).

\bibitem{Ve92}
A.P. Veselov, {\em Growth and Integrability in the Dynamics of Mappings}.
\newblock Comm.\ Math.\ Phys. {\bf 145} (1992), pp. 181--193.

\bibitem{Ar90}
V.I. Arnold, {\em Dynamics of complexity of intersections}.
\newblock Bol. Soc. Bras. Mat. {\bf 21} (1990), pp. 1--10.

\bibitem{GrRaPa91}
B.~Grammaticos, A.~Ramani, and V.~Papageorgiou, {\em Do integrable mappings
  have the {P}ainlev\'e property?}
\newblock Phys.\ Rev.\ Lett. {\bf 67} (1991), pp. 1825--1827.

\bibitem{RaGrHi91b}
A.~Ramani, B.~Grammaticos, and J.~Hietarinta, {\em Discrete versions of the
  {P}ainlev\'e equations}.
\newblock Phys.\ Rev.\ Lett. {\bf 67} (1991), pp. 1829--1832.

\bibitem{GrNiRa96}
B.~Grammaticos, F.W. Nijhoff, and A.~Ramani.
\newblock Discrete painlev\'e equations.
\newblock In R.~Conte, editor, {\em The Painlev\'e property, one century
  later},  (1996).
\newblock Carg\`ese proceedings, to appear in CRM proceedings and lecture
  notes.

\bibitem{HiVi97b}
J.~Hietarinta and C.-M. Viallet.
\newblock {\em Discrete Painlev\'e and singularity confinement in projective
  space}.
\newblock proceeding Bruxelles july 1997 meeting on "Integrability and chaos in
  discrete systems",  (1997).

\bibitem{Mu95}
D.~Mumford.
\newblock {\em Algebraic geometry, Vol.\ 1: Complex projective varieties}.
\newblock Springer-Verlag,  (1995).

\bibitem{He76}
M.~H\'enon, {\em A two-dimensional mapping with a strange attractor}.
\newblock Comm.\ Math.\ Phys. {\bf 50} (1976), pp. 69--77.

\bibitem{BeMaVi91c}
M.P. Bellon, J-M. Maillard, and C-M. Viallet, {\em Infinite Discrete Symmetry
  Group for the {Y}ang-{B}axter Equations: Spin models}.
\newblock Phys.\ Lett. {\bf A 157} (1991), pp. 343--353.

\bibitem{AbAnBoMa98}
S.~Boukraa N.~Abenkova, J-C. Angl\`es~d'Auriac and J-M. Maillard.
\newblock In preparation,  (1998).

\end{thebibliography}
\end{document}